\documentclass[letter]{aa}
\usepackage{graphicx}
\usepackage[varg]{txfonts}
\usepackage{longtable}
\usepackage{multicol}
\usepackage{color}
\usepackage{natbib}
\usepackage{upgreek}
\bibpunct{(}{)}{;}{a}{}{,} 

\newcommand{\frb}{FRB\,150418}
\newcommand{\wise}{WISE\,J0716$-$19}
\newcommand{\ujb}{$\upmu\mathrm{Jy} \, \mathrm{beam}^{-1}$}
\newcommand{\uj}{$\upmu\mathrm{Jy}$}

\def\ltsima{$\; \buildrel < \over \sim \;$}
\def\simlt{\lower.5ex\hbox{\ltsima}} 
\def\gtsima{$\; \buildrel > \over \sim \;$}
\def\simgt{\lower.5ex\hbox{\gtsima}} 
\defcitealias{Keane2016}{K16}
\defcitealias{Williams2016a}{WB16a}\defcitealias{Williams2016b}{WB16b}
\defcitealias{Bassa2016a}{B16}
\defcitealias{Vedantham2016}{V16}
\begin{document}

   \title{On the nature of \frb:\ clues from European VLBI Network and e-MERLIN observations}

   \author{M.\ Giroletti\inst{1}\fnmsep\thanks{Email: giroletti@ira.inaf.it}, B.\ Marcote\inst{2}, M.\ Garrett\inst{3},  Z.\ Paragi\inst{2}, J.\ Yang\inst{4}, K.\ Hada\inst{5}, T.\,W.\,B.\ Muxlow\inst{6}, C.\,C.\ Cheung\inst{7}
           }

   \institute{INAF Istituto di Radioastronomia, via Gobetti 101, 40129 Bologna, Italy \and
Joint Institute for VLBI ERIC, Postbus 2, 7990 AA Dwingeloo, The Netherlands \and
Netherlands Institute for Radio Astronomy (ASTRON), Postbus 2, NL-7990 AA Dwingeloo, The Netherlands \and
Department of Earth and Space Sciences, Chalmers University of Techn., Onsala Space Observatory, SE-43992 Onsala, Sweden \and
Mizusawa VLBI Observatory, National Astronomical Observatory of Japan, Osawa, Mitaka, Tokyo 181-8588, Japan \and
Jodrell Bank Centre for Astrophysics/e-MERLIN, The University of Manchester, M13 9PL, UK \and
   Space Science Division, Naval Research Laboratory, Washington, DC 20375-5352, USA }

\date{Received ; accepted }

  \abstract
   {}
   {We investigate the nature of the compact, and possibly variable nuclear radio source in the centre of \wise, the proposed host galaxy of fast radio burst, \frb.}
   {We observed \wise\ at 5.0 GHz with the European VLBI Network four times between 2016 March 16 and June 2. At three epochs, we simultaneously observed the source with e-MERLIN at the same frequency.}
   {We detected a compact source in the EVN data in each epoch with a significance up to $\sim 8\sigma$. The four epochs yielded consistent results within their uncertainties, for both peak surface intensity and positions. The mean values for these quantities are $I_\mathrm{peak}=(115\pm9)$ \ujb\ and  r.a.\ = 07$^{\rm h}$ 16$^{\rm m}$ 34.55496(7)$^{\rm s}$, dec.\ = $-19^\circ$ 00\arcmin\ 39.4754(8)\arcsec, respectively. The e-MERLIN data provided $\sim 3-5\sigma$ detections, at a position consistent with those of the EVN data. The presence of emission on angular scales intermediate between the EVN and e-MERLIN is consistent with being null. The brightness temperature of the EVN core is $T_{\rm b} \gtrsim 10^{8.5} {\rm K}$, close to the value required by Akiyama \& Johnson (2016) to explain the radio properties  of \wise\ in terms of interstellar induced short-term variability.}
   {Our observations provide direct, independent evidence of the existence of a nuclear compact source in \wise, a physical scenario with no evident connection with \frb. However, the EVN data do not show indication of the variability observed with the VLA.}

\clearpage

   \keywords{galaxies: active -- galaxies: individual: WISE J071634.59$-$190039.2 -- radio continuum: galaxies -- scattering
               }

  \authorrunning{M.\ Giroletti et al.}
  \titlerunning{Is \frb\ localised in \wise? }

   \maketitle
%

\section{Introduction}

Fast radio bursts (FRBs) are transient episodes characterised by short (sub-ms) duration and large dispersion measure (DM). After the initial discovery by \citet{Lorimer2007}, several new such events have been discovered \citep[e.g.,][]{Thornton2013,Champion2016}, triggering debate about their nature. It is possible that they are due to young, highly magnetized neutron stars, as suggested for the repeating FRB\,121102 \citep{Spitler2016}, or to cataclysmic events.

Both Galactic and extragalactic origins have been proposed. An extragalactic origin is preferred based on the large DM, $\simgt (0.5-1) \times 10^{3}$ cm$^{-3}$ pc typically found; however, only a precise localisation and a measurement of the redshift could be conclusive. For this reason, the reported localisation of \frb\ to the elliptical galaxy, WISE J071634.59$-$190039.2 (hereafter, \wise) by \citet[][hereafter, \citetalias{Keane2016}]{Keane2016} has attracted interest in the community.  The precise redshift determination for this FRB has immediate implications for the system's energetics, thus possible FRB progenitors \citep{Liu2016,Zhang2016}, as well as applications to probe fundamental physics \citep{Bonetti2016,Tingay2016}.

\begin{table*}
\centering
\caption{Log of observations, image parameters and model fit results. \label{t.log}}
\begin{tabular}{lcccccccccc}
\hline
\hline
\multicolumn{2}{c}{Epoch} & \multicolumn{4}{c}{EVN data} &  \multicolumn{4}{c}{e-MERLIN data} & \\
Date in & & HPBW & $I_{\rm peak}$ & $I_{\rm noise} $& $S_{\rm 5.0, JMFIT}$ & HPBW & $I_{\rm peak}$ & $I_{\rm noise} $& $S_{\rm 5.0, JMFIT}$ & $\Delta S_{\rm 5.0}$\\
2016 & MJD & (mas $\times$ mas, $^\circ$) & \multicolumn{2}{c}{(\ujb)} & (\uj) & (mas $\times$ mas, $^\circ$) & \multicolumn{2}{c}{(\ujb)} & (\uj)\\
\multicolumn{1}{c}{(1)} & \multicolumn{1}{c}{(2)} & \multicolumn{1}{c}{(3)} & \multicolumn{1}{c}{(4)} & \multicolumn{1}{c}{(5)} & \multicolumn{1}{c}{(6)} & \multicolumn{1}{c}{(7)} & \multicolumn{1}{c}{(8)} & (9) & (10) & (11) \\
\hline
March 16 & 57463.8 & $10.1 \times 6.2, 3.9$ & 123 & 18 & $125\pm22$ & \dots & \dots & \dots & \dots \\
May 10 & 57518.6 & $9.7 \times 6.1, 8.7$ & 113 & 14 & $137\pm20$ & $261\times25, 12$ & 169 & 55 & $176\pm58$ & $40\pm60$ \\
May 31 & 57539.6 & $10.9 \times 6.1, -7.5$ & 107 & 16 & $117\pm20$ & $231\times27, 11$ & 145 & 48 & $158\pm51$ & $40\pm55$ \\
June 2 & 57541.6 & $9.3 \times 5.3, 1.3$ & 133 & 20 & $125\pm32$ & $212 \times 28, 10$ & 246 & 52 & $264\pm59$ & $140\pm70$ \\
\hline
\end{tabular} 
\tablefoot{Cols.~(1, 2): observation date; Cols.~(3--6): EVN half-peak beam width (HPBW), naturally weighted image peak brightness and $1\sigma$ noise level, and results of a 2-d Gaussian fit to the image brightness distribution; Cols.~(7--10): same as Cols.~(3--6), for e-MERLIN data; Col.~(11): flux density difference between e-MERLIN and EVN data.}
\end{table*}

The proposed identification of \frb\ with \wise\ was based on the prompt detection (beginning 2 hrs after the FRB discovery) with the Australia Telescope Compact Array (ATCA) of a fading radio source within one beam of the 21-cm Parkes multi-beam receiver. Optical photometric and spectroscopic follow-up observations with the Subaru telescope identified an elliptical galaxy at $z=0.492\pm0.008$ consistent with the radio source within the $\sim1\arcsec$ positional uncertainty. In more detail, the radio transient emission was observed in only the first two epochs of ATCA follow-up separated by 6-days (flux densities $\sim 0.2$ mJy at 5.5 GHz) with three subsequent detections of essentially steady emission ($\sim 0.1$ mJy at 5.5 GHz) attributed to the emission from the host galaxy. 

This association of the prompt ms-duration emission from the FRB with the variable ATCA source has been questioned. \citet[][hereafter \citetalias{Williams2016a}]{Williams2016a} argued instead that \wise\ is consistent with being a random active galactic nucleus (AGN) found within the Parkes beam based on the known rate of variable (rather than transient) radio sources, that the steady radio emission component implies a large luminosity more typical of an AGN, and that the radio light curve is inconsistent with the evolution of a standard afterglow.

Further criticism of the proposed association came from the results of a Karl G.\ Jansky Very Large Array (VLA) observing campaign almost a year after the FRB: in 10 total observations at 5.5 and 7.5 GHz spanning 35 days, \citet{Williams2016a,Williams2016b} found variable radio emission at an enhanced level with respect to the previously observed steady $\sim 0.1$ mJy source. \citet[][hereafter \citetalias{Vedantham2016}]{Vedantham2016} also reported a single epoch multi-frequency VLA observation over the 1-18 GHz frequency range that showed a flat spectrum radio source consistent with an AGN. Finally, numerical simulations by \citet{Akiyama2016} indicate that the reported light curve is consistent with scintillating radio emission from an AGN core with $T_\mathrm{b} \simgt 10^9 \, \mathrm{K}$.

A final confirmation of the AGN scenario, plus a relevant contribution from refractive interstellar scintillation, can be obtained from high angular resolution Very Long Baseline Interferometry (VLBI) observations. In this Letter, we thus report on the results of European VLBI Network (EVN) e-MERLIN observations of \wise. In the following, we describe the observations in Sect.\ 2, present the results in Sect.~3, and discuss them in Sect.~4.

\section{Observations and data reduction\label{s.observations}}

We observed \wise\ four times between 2016 March 16 and June 2 (Table~\ref{t.log}) with a subset of the EVN. The participating stations were Effelsberg, Hartebeesthoek, Jodrell Bank (Mark2), Medicina, Noto, Onsala, Torun, Yebes, and a single Westerbork telescope. We observed at 5.0 GHz, with eight 16-MHz-wide baseband channels, in dual polarization, and with 2-bit sampling. The data were electronically transferred over fibre links to the SFXC correlator at JIVE, where they were correlated in real time with the so-called e-VLBI technique. 

We carried out all observations in phase-reference mode, with 2.5\,min scans on the target source bracketed by 1.5\,min scans on the nearby ($0.9^\circ$ offset) calibrator J0718--1813. Each observation lasted for $\sim 5.5$ hours, with on-source time of $\sim 2.4$ hours. We calibrated visibility amplitudes based on the a-priori gain curves and measured system temperatures at each station. Parallactic angle corrections were applied and we determined instrumental single band delays using a scan on a strong calibrator. We then determined phase, rates, and residual delays for the phase calibrator. Since the calibrator has a double component structure, we imaged it with hybrid mapping procedure, and then repeated the fringe fitting process using the obtained image as the input model. The resultant solutions were applied back to the phase reference source, the target, and the additional check source J0712--1847. Bandpass solutions were then determined combining all the data for the calibrator. Finally, we carried out one cycle of phase-only and one of phase-and-amplitude self-calibration for the phase reference source, and transferred the solutions to the target. A parallel analysis of the check source based on either direct fringe fitting of its visibility data or phase self-calibration indicated that coherence losses affected the detected ranging between $20\%$ and $40\%$ of the real flux density; this is not surprising given the low elevation of the target.

Initially, we imaged the first epoch data over a large field of $4\arcsec \times 4\arcsec$, centred on the WISE coordinates, r.a.\ = 07$^{\rm h}$ 16$^{\rm m}$ 34.59$^{\rm s}$, dec.\ = $-19^\circ$ 00\arcmin 39.2\arcsec. The overall rms noise was about 25 \ujb\ \citep{Marcote2016}. Following the report of the VLBA and e-MERLIN localisation by \citet[][hereafter \citetalias{Bassa2016a}]{Bassa2016a}, we imaged a smaller field around their preliminary VLBA-measured position. The local noise for an image produced with the AIPS task {\tt IMAGR} using {\tt ROBUST = 5} is 18 \ujb. In the following epochs, we reached similar or better noise values, except for the last epoch in which the most sensitive telescope (Effelsberg) did not provide data for about a half of the observation run.

In strict simultaneity with the latter three EVN epochs (same start and end times), we observed the source with e-MERLIN, using six, five, and five stations in each experiment. We observed at 5.0 GHz, with four 128-MHz-wide channels, in dual polarisation. The same phase reference source was used as in the EVN run. The maximum elevation of the source was $18^\circ$, which resulted in an elongated restoring beam (axial ratio $\sim10$, in p.a.\ = $\sim 10^\circ$). Detailed information is reported in Table~\ref{t.log}.

\begin{table*}
\centering
\caption{Normalised EVN peak brightness of target and calibrator in time bins. \label{t.bins}}
\begin{tabular}{lcccccc}
\hline
\hline
& \multicolumn{3}{c}{\wise} &  \multicolumn{3}{c}{J1718--1813} \\
Epoch & $S_1/\langle S \rangle$ & $S_2/\langle S \rangle$ & $S_3/\langle S \rangle$ & $S_1/\langle S \rangle$ & $S_2/\langle S \rangle$ & $S_3/\langle S \rangle$ \\
\hline
1 & $	1.13	\pm	0.25	$ & $	1.14	\pm	0.25	$ & $	0.74	\pm	0.23	$ & $	0.93	\pm	0.09	$ & $	0.99	\pm	0.10	$ & $	1.08	\pm	0.11	$ \\
2 & $	0.90	\pm	0.22	$ & $	0.93	\pm	0.22	$ & $	1.16	\pm	0.23	$ & $	0.97	\pm	0.10	$ & $	1.00	\pm	0.10	$ & $	1.04	\pm	0.10	$ \\
3 & $	1.32	\pm	0.30	$ & $	0.95	\pm	0.29	$ & $	0.73	\pm	0.28	$ & $	0.92	\pm	0.09	$ & $	1.00	\pm	0.10	$ & $	1.08	\pm	0.11	$ \\
4 & $	0.88	\pm	0.26	$ & $	0.53	\pm	0.25	$ & $	1.60	\pm	0.29	$ & $	0.92	\pm	0.09	$ & $	1.02	\pm	0.10	$ & $	1.05	\pm	0.11	$ \\
\hline
\end{tabular} 
\end{table*}

\begin{figure*}
\center \includegraphics[width=14.0cm]{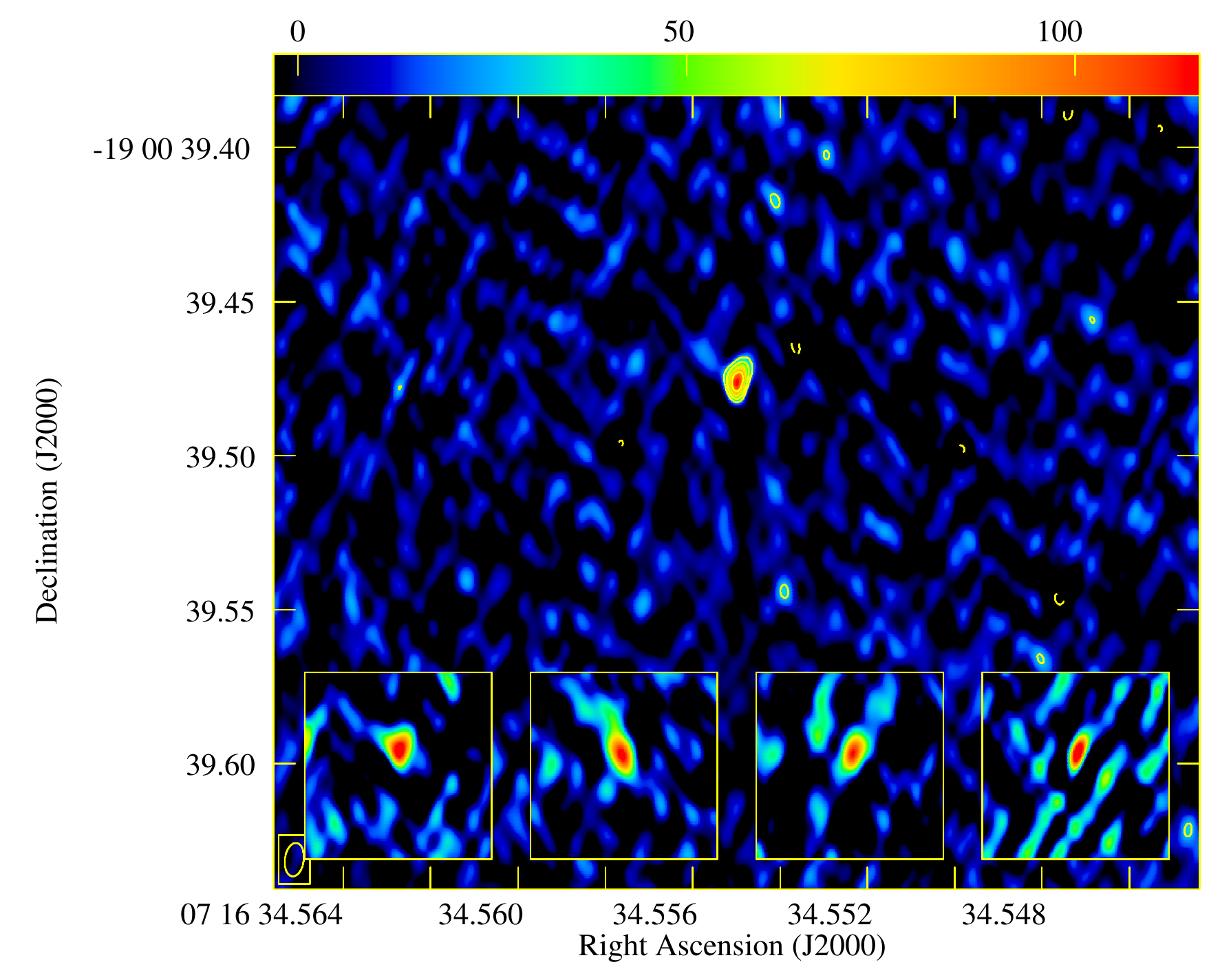}
\caption{EVN 5.0 GHz images of \wise. Main panel: average image obtained combining all four observations; contours are traced at $\pm 3, 5, 10 \times$ the image local rms noise of 8.9 \ujb; the peak surface brightness is 115 \ujb; the restoring beam is shown in the bottom left corner and it is 10.9 mas $\times$ 6.1 mas in p.a.\ = $-7^\circ$; the colour scale indicates surface brightness between $-3.0$ and 115 \ujb. Insets: central part of the field, as obtained from each individual dataset; pixel size and colour scale are the same as in the main panel, for ease of comparison. \label{f.1}}
\end{figure*}

\section{Results\label{s.results}}

In Fig.~\ref{f.1}, we show our EVN 5.0 GHz images around the position of the VLBA and e-MERLIN detections reported by \citetalias{Bassa2016a}. The main image shows a $0.3\arcsec \times 0.25\arcsec$ field-of-view based on averaging the images from all epochs. The insets show 60 mas $\times$ 60 mas image stamps of the central region from the four individual epochs. 

In each of the individual epochs, the source is detected with significances above $6\sigma$. In Cols.\ 4, 5, and 6 of Table~\ref{t.log}, we report the image peak brightness, the noise, and the component flux density measured with AIPS task {\tt JMFIT}. The associated uncertainties were calculated as the quadratic sum of a $1\sigma$ r.m.s.\ statistical contribution and a 10\% absolute calibration uncertainty; this provides the uncertainty on the relative calibration from epoch to epoch, while the overall scaling due to coherence losses remains unaccounted for. Within these uncertainties, the source flux density is consistent with being constant among epochs; the best fit coordinates are also consistent to 1/10 of the restoring beam, or less. 

In the stacked EVN 5.0 GHz image, the peak brightness and the rms noise are $I_{\rm peak}=115$ \ujb\ and $I_{\rm noise}=8.9$ \ujb, respectively, amounting to an overall $>12\sigma$ significance detection. Given the lack of significant variability among epochs, we determined the source parameters from a fit to the mean image, giving $S_{\rm 5.0, JMFIT} = (120\pm15)$ \uj, r.a.\ = 07$^{\rm h}$ 16$^{\rm m}$ 34.55496(7)$^{\rm s}$, dec.\ = $-19^\circ$ 00\arcmin\ 39.4754(8)\arcsec\ (the digits in parenthesis indicate the standard deviation of the measurements of the four datasets)\footnote{This position is also in agreement with that reported by \citet[][published during the revision process of this letter]{Bassa2016b}}. At the luminosity distance of \wise\ ($d_L=2.81$ Gpc), the corresponding average monochromatic luminosity is $L_{\rm 5.0} = (1.13\pm 0.15) \times10^{23}$\,W\,Hz$^{-1}$.

Our array has a large gap in the $(u,v)$-plane between European baselines and baselines to South Africa; moreover, the phase reference source has a double structure and the phase solutions to Hartebeesthoek are less stable. For these reasons, we determined the above parameters on images based only on the European baseline data. However, the source is still significantly detected also in images with the full array; it is consistent with being unresolved, with a peak brightness $I_{\rm peak}=118$ \ujb\ and a deconvolved size $\theta < 0.34$ mas. We also carried out model-fits using Difmap, fitting a circular Gaussian component to the visibility data. The fit provides a major axis size of 0.18 mas. Taking the visibility plane model-fit size as an upper limit, we estimate a brightness temperature of $T_\mathrm{B} \ge 10^{8.5}$ K.

\begin{figure*}
\center \includegraphics[width=\textwidth]{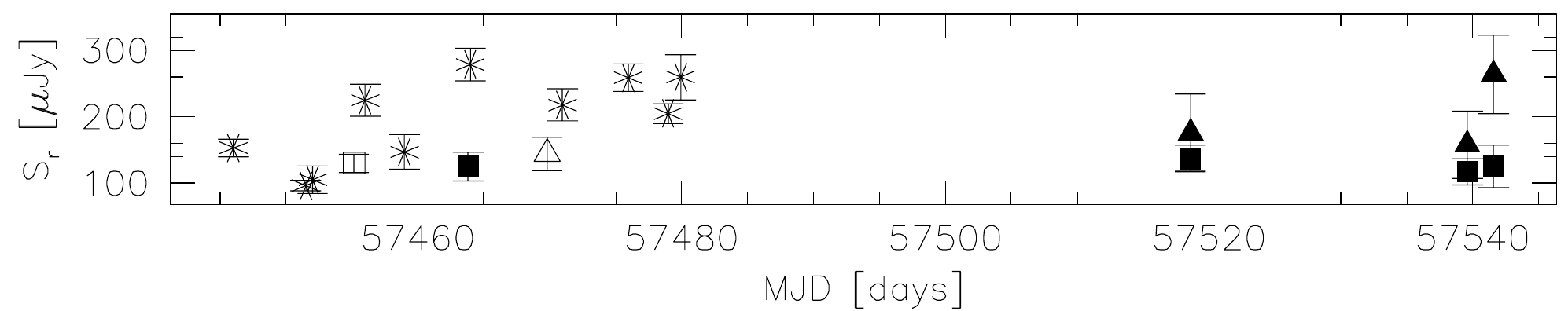}
\caption{Light curve of \wise. Stars: 5.5 GHz VLA data \citepalias{Vedantham2016,Williams2016a,Williams2016b}; squares: 5.0 GHz VLBI data; triangles: 5.0 GHz e-MERLIN data. The filled symbols are from this work, empty ones from \citetalias{Bassa2016a}.
\label{f.lc}}
\end{figure*}
No significant variability is present from one epoch to the other. We explored the presence of variability on shorter time scales dividing each observation in three bins of $\sim2$-hr duration each. The values of the target and of the calibrator peak brightness in each bin are reported in Table~\ref{t.bins}, normalised to the mean of each epoch. The source is generally detected in every subsets of dataset, albeit at lower significance (notice that the relative errors are comparatively large, as the noise in each bin is $\sim \sqrt{3} \times$ higher than in the full dataset). In 8/12 cases, the peak brightness is consistent with the mean value. In the last epoch, two bins are about $2\sigma$ away from the mean. The significance of these variations is difficult to establish because the $(u,v)$-plane coverage is different in every bin, as was the elevation of the source. The calibrator peak brightness is more stable, yet it generally increases in the second and third subsets of each observation, as the restoring beam rotates closer to the main p.a.\ of the double structure of the source. While this indicates that no systematic effects due to calibration are present, it also shows the difficulty in establishing variability on short time scales with the present data.

Finally, we detect a point-like source  in every e-MERLIN  observation, at a significance level of just above $3\sigma$. The coordinates are consistent with those obtained with the EVN, yet less well determined due to the elongated beam. The peak brightness in two epochs is also consistent with that detected by the EVN. In the final epoch, the peak is higher, although we caution that the weak signal-to-noise ratio and the low elevation indicate that this might not be a highly significant discrepancy.

\section{Discussion}

The presence of an AGN within \wise\ was already implicit in the results presented by \citetalias{Keane2016}. They reported an upper limit to the H$\upalpha$ luminosity associated with a star-formation rate of $\le 0.2 M_\odot \, {\rm yr}^{-1}$. Based on \citet{Condon1992}, this value corresponds to a radio luminosity $L_{\rm 5.0\, GHz} \le 10^{21}$\,W\,Hz$^{-1}$, about two orders of magnitude lower than that observed at the ATCA quiescence level. 
Our observations now provide a firm proof of the presence of a compact radio source at the centre of \wise, with a bolometric radio luminosity of $\nu L_\nu = 5.6 \times 10^{39}$ erg\,s$^{-1}$ and not variable, within the uncertainties. At first sight, this result supports the association proposed by \citetalias{Keane2016} between \frb\ and the subsequent episode of variable radio emission. 

However, the VLA data \citepalias[][angular resolution of $\sim 8\arcsec \times 3\arcsec$ at 5.0 GHz]{Williams2016a} indicate that variability is still present in the source more than one year after \frb: in Fig.~\ref{f.lc}, we show the light curve at 5.0 and 5.5 GHz over the time range 2016 February 27 to June 2, obtained with the VLA, EVN, VLBA, and e-MERLIN. The VLA measurements are generally higher and more variable than the EVN ones: the mean flux density and the variability index are $\langle S_{\rm VLA}\rangle = 195$ \uj, $V_{\rm VLA} = 0.49$ for the VLA, and $\langle S_{\rm VLBI}\rangle = 127$ \uj, $V_{\rm VLBI} = 0.08$ for the VLBI data (i.e.\ both EVN and VLBA). We exclude that this discrepancy is due to the presence of a secondary variable component in addition to the core imaged with EVN and e-MERLIN. Causality forbids variability on $\sim$ day time scales from diffuse emission resolved out by the EVN and e-MERLIN baselines (scale of $> 0.2\arcsec \sim 1.3$ kpc), which would also be inconsistent with the  star formation rates determined by \citetalias{Keane2016}. The EVN data themselves do not show evidence for any secondary compact component either, in particular around MJD 57463, when the EVN and VLA data are nearly simultaneous, and differ by 150 \ujb; no non-nuclear sources are known to reach such a large luminosity.

This requires us to explore the time, rather than the spatial, domain. It is possible, although unlikely, that the discrepancy is a chance coincidence: a K-S test on the distribution of the VLA and VLBI flux densities provides a probability that the two are drawn from the same distribution as low as 0.011. There is one significant factor to take into account: due to the different sensitivity of the two instruments, VLA data are obtained on much shorter time scales (typically, 30 minutes) than the EVN's (many hours). We can thus hypothesise that the parsec-scale source varies on short ($<$ hr) time scales, so that the VLA-based light curve resolves the variations, while they are averaged out by the longer EVN observations. 

The above scenario does also present some challenges. Intrinsic sub-hour time-scale variability from AGNs requires extremely large brightness temperature, exceeding the inverse Compton catastrophe limit. On the other hand, \wise\ is located at low Galactic latitude ($b=-3^\circ.2$), indicating that radio waves are subject to significant refractive scintillation in the ionized interstellar medium of the Milky Way. \citet{Akiyama2016} have argued that few-day time-scale variability of \wise\ could indeed be extrinsic, if  the source has a  $T_{\rm b} \gtrsim 10^9 {\rm K}$, which is consistent with our result. Scintillation has so far been studied mostly in blazars and little is known about the variability properties of weak sources; however, very rapid variations in \wise\ would at least be in agreement with the trend of increased variability found for lower flux density sources \citep{Lovell2008}.

\begin{acknowledgements}
We thank J.~Y.~Koay for useful discussions.
The European VLBI Network is a joint facility of independent European, African, Asian, and North American radio astronomy institutes. Scientific results from data presented in this publication are derived from the following EVN project codes: RG008A-D. e-MERLIN is a National Facility operated by the University of Manchester at Jodrell Bank Observatory on behalf of STFC.
Work by C.C.C. at NRL is supported in part by NASA DPR S-15633-Y.

\end{acknowledgements}

\end{document}